\documentclass[12pt,preprint2]{aastex}  
\accepted{}
\newcommand {\tighttable}{\def\baselinestretch{1.0}}
\newcommand {\flx}{${\rm erg\, cm}^{-2}\,{\rm s}^{-1}\,{\rm Hz}^{-1}{\rm Sr}^{-1}$}
\newcommand {\ergscmA}{${\rm erg\,s}^{-1}\,{\rm cm}^{-2}\,{\rm \AA}^{-1}$}
\newcommand {\cl}{$\rm cm^{-2}$}
\newcommand {\kms}{${\rm km\,s}^{-1}$}
\newcommand {\lya}{Ly$\alpha$}
\newcommand {\hb}{H$\beta$}
\newcommand {\h}{\ion{H}{1}}

\newcommand {\he}{\ion{He}{2}}
\newcommand {\oi}{\ion{O}{1} $\lambda 1304$}
\newcommand {\ovi}{\ion{O}{6} $\lambda 1035$}
\newcommand {\siiv}{\ion{Si}{4} $\lambda 1402$}
\newcommand {\cii}{\ion{C}{2} $\lambda 1335$}
\newcommand {\ciii}{\ion{C}{3}]}
\newcommand {\civ}{\ion{C}{4} $\lambda 1549$}
\newcommand {\nv}{\ion{N}{5} $\lambda 1240$}
\newcommand {\mgii}{\ion{Mg}{2} $\lambda 2799$}
\usepackage{multirow}
\shorttitle{Quasar Age}
\shortauthors{Zheng} 
\begin{document}
\title{Spectral Signatures of Quasar Ages at ${\emph z \approx 3}$ \altaffilmark{1,2}}
\author{Wei Zheng\altaffilmark{3}} 
\altaffiltext{1}{{\it The Astrophysical Journal}, {\bf 892}, 139 https://doi.org/10.3847/1538-4357/ab7b6f}
\altaffiltext{2}{In memory of E. Margaret Burbidge (1919-2020)}
\altaffiltext{3}{Department of Physics and Astronomy, Johns Hopkins University, 3701 San Martin Dr., Baltimore, MD 21218} 
\begin{abstract}
Insight into quasar ages may be obtained from the proximity effect, but so far only 
in a limited number of bright quasars. Based on $\sim 2600$ SDSS quasar spectra at $2.5 \leq z \leq 3.5$, 
a search for spectral voids between \lya\ forest lines finds
proximity zones over a wide range of radial distances. The majority of zone sizes are less than 5 Mpc, 
with their numbers decreasing exponentially towards larger distances.
After normalization by luminosities, the zone sizes 
are distributed with an e-folding scale of 0.64 as compared with 
 the anticipated values. A group of quasars are selected for their large proximity zones of $\gtrsim 10$ Mpc. 
Their composite spectrum displays strong narrow cores and large equivalent widths in \lya\ 
and other major UV emission lines.
If the proximity zones along lines of sight are indicative of quasar ages, 
these features may be the signatures of old quasars.
Another group of quasars are selected as they show no proximity zone and exhibit intrinsic absorption lines at $z_{ab} > z_{em}$. 
They are likely young quasars and exhibit weaker narrow emission-line components.
The significant difference of spectral features between the two groups 
may reflect an evolution pattern over quasars' lifetimes. 
\end{abstract}

\section{INTRODUCTION}\label{sec_intr}

What is a quasar's lifetime? How do quasars evolve over their lifetimes? To answer these questions regarding supermassive black holes, it is essential to find clues to the 
age of quasars. 
However, such  time tags are scarce as quasar spectra, unlike those of stars and galaxies, 
consist of power-law segments that extend to high frequencies  
\citep{elvis,shang}. While the UV bumps are believed to be related to the accretion disks, the estimates of 
their temperatures are difficult \citep{blackburne}.
Various methods have been tried to probe quasar ages, but the results are uncertain and 
sometimes inconsistent, 
with a possible range between $1-100$ Myr \citep{martini,kirkman,dipompeo}. 

Perhaps the best trace for a quasar's age is in its vicinity.
The enormous UV radiation of quasars produces cosmic bubbles of high ionization that expand and eventually overlap. 
The accumulated ionizing radiation field, commonly referred to as the 
metagalactic UV background (UVB) radiation, completed the reionization 
of the intergalactic medium \citep[IGM; ][]{meiksin,mcquinn}
at $z \approx 3$.  
The trace of high-ionization zones around quasars can be found in their spectra as the 
line-of-sight proximity effect \citep{carswell,murdoch,bdo}. In the hydrogen \lya\ forest lines at
$z \approx 3$, the effect is observed as a decrease in the number of forest absorption lines. 
In the helium \lya\ spectra at $z \approx 3$, 
the opacity is large enough that a continuous proximity profile may be present \citep{zheng95}. The helium proximity profiles in a few dozens of quasars \citep{zheng15,khrykin,khrykin19,zheng19}
have been used to estimate quasar ages. 

Early studies of the proximity effect were carried out with high-resolution optical spectra where a large number of 
absorption lines could be identified \citep{carswell87,giallongo,lu96,cooke}. Later studies
\citep{bechtold,scott,liske,guimaraes} often used spectra of medium resolution of $\sim 4000$
to find a deficiency of absorption lines near the quasar redshifts and derive the UVB
intensity. \cite{dall} used
high signal-to-noise (S/N), low-resolution 
($R\approx 800$) spectra to carry out such a study. The largest sample size 
among these studies is 45 \citep{guimaraes}. 

The proximity effect can also be studied along the transverse direction if a bright foreground quasar is present 
near the line of sight towards another distant quasar \citep{dobrzycki}. 
Most \h\ studies in the optical band have not yielded significant detections \citep{liske,kirkman,lau}, but 
\he\ studies have made several estimates of quasar ages of up to $\sim 34$ Myr \citep{jakobsen,syphers14,schmidt17,schmidt18}.

The Sloan Digital Sky Survey \citep[SDSS;][]{york}, its follow-up SDSS-II Supernova Survey
\citep{friedman}, SDSS-III/BOSS \citep{eisenstein}, and SDSS-IV/eBOSS \citep{eboss}
have revolutionized the way we study quasars. Over the last two decades, the number of known 
quasars has increased dramatically from several thousand to 
approximately half a million \citep{q7,q10,q12,q14}.
This paper describes an analysis of medium-resolution SDSS spectra in search 
of the proximity effect. It is known that some quasars display
 small proximity zones: for \he\ zones at $z\approx 3$ \citep{shull,zheng15,khrykin} and \h\ zones 
at $z\approx 6$ \citep{eilers18}. Does such a trend exist for \h\ zones at $z\approx 3$?
The significant sample size enables one to derive the distribution of proximity-zone sizes and to 
identify groups of both old and young quasars. 

\section{DATA ANALYSES}\label{sec_data}

 Quasar spectra at $2.5 \leq z \leq 3.5$ and with a limiting magnitude of 19 in the 
$i'$ band were 
retrieved from the Data Releases 7, 10, and 12 \citep[DR,][ respectively]{dr7,dr10,dr12}.
Figure \ref{fig-sam} displays the distribution of these 3298 quasars. 
SDSS spectra cover a wavelength range $\sim 3600 - 10500$~\AA, at a resolution of $\sim 2000$. 
The data at the red and blue ends are of low S/N ratio, therefore, the 
nominal spectra range is $\sim 3800 - 9200$~\AA.
At redshifts of 2.5 and higher, the \lya\ emission moves into the region of $\lambda > 4255$~\AA, 
securing a spectral window $\Delta\lambda \gtrsim 450$~\AA\ with reasonable S/N ratios for 
the study of forest 
absorption. The upper limit of $z=3.5$ is set to secure the \ciii\ $\lambda 1909$ (\ciii\ hereafter)
 emission line well below 9000~\AA\ to avoid significant night-sky lines.
The average redshift of this sample is $z = 2.84 \pm 0.26$  with a median of 2.78.
The spectra were processed in units of vacuum wavelength. 
When multiple spectra exist for a given quasar, they  were merged with the weights of S/N ratios.  
All the wavelengths and equivalent widths (EW) are in the rest frame, and all distances as proper 
distances hereafter. At $z\approx 3$, the distance scale is approximately 0.8 Mpc per \AA. 

\subsection{Simulations of \lya\ Forest Lines}
\label{sec_sim}

To check the feasibility of finding the proximity effect with SDSS spectra, \lya\ forest-line spectra 
were simulated at $z \approx 3$.
Following an empirical power-law distribution of column densities $dn/dN \propto N^{-1.5}$ \citep{tytler},
absorbers were generated between column densities of $\log N = 11 - 17$~\cl\ and $\lambda = 1050 - 1216$~\AA. 
The wavelengths of these absorbers
are random but follow a distribution of $dn/dz \propto (1+z)^{2.5}$ \citep{janknecht}.
For every absorber, a Voigt profile of Doppler parameter $b=30$~\kms\ was produced on a 
flat continuum over a grid of 0.01~\AA\ scale. 
Over the range of $1050 - 1170$~\AA, there are $\sim 2700$ lines, among which $\sim 67$ are at 
$\log N > 13.5$~\cl.
The spectra were then binned to match the SDSS resolution with added noise at an S/N ratio of 20. 
The left panels of Figure \ref{fig-sim} display an example of simulated data: the column densities of 
absorbers, the normalized forest-line spectrum and the RMS-fluctuation spectrum (absolute values of the 
flux difference between adjacent pixels). The effective optical depth produced by all \lya\ absorption may be
expressed as $\tau = 0.0021\ (1+z)^{3.7}$ \citep{meiksin06}, and the simulations found that 
weak absorbers of column density $\log N \lesssim 13.4$~\cl\ contribute approximately a half of this opacity.

The proximity effect is commonly characterized by the ratio of the local
quasar flux to that of the UVB, $\omega = F_Q/(4\pi J_\nu)$, where 
$J_\nu$ is the angle-averaged specific intensity 
\citep{bdo}. For the UVB intensity $\log J_\nu \simeq 
-21.5$ \flx\  \citep{dall2,bolton05}, 
the proximity zones of $\omega \approx 1$ are moderate, on the order of 
5 Mpc. In the right panels of Figure~\ref{fig-sim}, the properties of a large proximity zone with $R_{\omega=1}=20$~\AA\ ($\approx 16$~Mpc) are plotted. 

Most \lya\ absorption lines that can be identified in SDSS spectra are saturated. 
Their strengths are therefore insensitive to the proximity effect, and few vanish in 
the quasar vicinity. A significant effect is on the weak components at 
$\log N \lesssim 13.8$~\cl\ (EW $\lesssim 0.1$~\AA). They are located 
on the linear part of the curve of 
growth, and their strengths are proportional to changing column densities.
While these weak absorption features cannot be individually identified at medium resolutions, 
their combined proximity effects are spectral voids between strong absorption lines, as shown in 
the right panels of Figure~\ref{fig-sim}. Given the limited resolution and S/N ratios of SDSS spectra, the spectral 
voids in the forest-line 
region are used in this study as a signature of the proximity effect.

\subsection{RMS Fluctuation Spectra}
\label{sec_prep}

In a process of flux normalization, absorption lines in the SDSS spectra were first 
identified using local flux troughs. 
By making the flux differences between adjacent pixels in a 
spectrum, an array of first-order derivative was made, and then was 
a second-order derivative by repeating the same method to the latter.
After a local trough was identified 
from the second-order derivative, a ``climbing" process started 
towards both the longer and shorter wavelengths until nearby ``plateaus" were found, which
marked the endpoints of this absorption feature.
The EW and statistical significance were calculated, and the corresponding wavelength
window of this absorption feature was flagged out. The weakest features that could be identified were of EW~$ \approx 0.15$~\AA. 
The fitting task {\tt Specfit} \citep{specfit} was carried out over 
several wavelength ranges. 
The algorithm allows
a variety of input components with free parameters, including a power-law 
continuum, Gaussian and Lorentzian 
emission lines, Gaussian absorption lines, and user-supplied components.
During the analysis of a quasar spectrum, it first fit the wavelength 
region of $1216 - 1950$~\AA, excluding the spectral windows of identified absorption features. The components included a power-law continuum, Gaussian emission lines
for \lya, \nv, \civ, and \ciii. Secondly, it fit the wavelength region 
of $1150 - 1270$~\AA\ for improved normalization near and below the \lya\ wavelength. 
In the forest-line region, only a small fraction of pixels are free of absorption features. 
These high-flux points were included in the fitting windows.
At medium spectral resolutions, even these high points are below the flux extrapolation of a 
power law from longer wavelengths because of unresolved forest absorption lines. 
To mitigate this effect, a set of five ``userabs'' files with index keys were made in two columns: wavelength and optical depth. 
They represent a smooth attenuation component with $\tau = k\ 0.001\ (1+z)^{3.7} 
\ P$, where the coefficient 0.001 represents 
the effective optical depth of unresolved weak \lya\ forest lines 
(approximately a half of the total), the normalization factor $k$ between 0.0 and 1.0, 
and $P = (1+(R_{\omega=1}/R)^2)^{-1}$ the proximity term. 
Fixed characteristic distances $R_{\omega=1}$ of 2.5, 5, 10, 15, and 20
 \AA\ were assigned to these files, respectively, and radial distances
 from the quasar, $R$, were calculated as a function of wavelengths. 
This user-supplied component applied attenuation to all the pixels in the forest-line 
region. Its free parameters were the index key of input files ($1-5$) and the normalization scale
of optical depth, $k$. The fitting process yielded the best values of these two parameters, but their associated 
errors were large in many cases. 
The narrow components of emission lines were tied to that of \lya\ at 
FWHM  (Full Width at Half Maximum) $\leq 2500$ \kms,
and their broad counterparts tied to that of \lya\ at FWHM$ > 2500$ \kms. 
After these fitting steps, a normalized spectrum was produced for each quasar, with absorption-line windows flagged out.

A prominent difference between the forest-line region and the rest of a quasar spectrum is their RMS
characteristics. An RMS-fluctuation spectrum is made of the absolute values of flux 
difference between adjacent pixels in a normalized spectrum, after slight smoothing.
As shown in Figure~\ref{fig-sim}, the fluctuation properties are visibly different across the \lya\ wavelength. 
They provide a sensitive probe of spectral voids, and are 
largely unaffected by the flux normalization. 

A statistical analysis was carried out on RMS-fluctuation spectra in three regions: (A) the 
forest-line region between $1050-1175$~\AA, which is 
largely free of the proximity effect, 
(B) \lya\ emission-line region between $1216-1230$~\AA,
 and (C) two ``clean'' regions of $1260 - 1275$ and $1425 - 1500$~\AA. 
In the region A, all pixels were included, and in the region B and C, 
the wavelength windows with identified absorption features were excluded. 
The contrast parameter is defined as $\zeta = (\bar{A} - \bar{B})/\sigma(B)$, where $\bar{A}$ and $\bar{B}$ are mean fluxes in the 
A and B regions, respectively, and $\sigma(B)$ the standard deviation of fluxes in the B region. 
This ratio reflects the significance of the forest signature: a null $\zeta$ value
would imply no difference in fluctuation characters between regions A and B.
Figure \ref{fig-der} illustrates the fitting windows and the effect of 
parameters in a typical quasar.
In rare cases ($\sim 4.6\%$ of the whole sample), spectral region B was 
affected by many absorption features, and a better contrast ratio between regions A 
and C was used.
Since the potential proximity zone and regions B are at higher fluxes than the continuum level, 
the average and standard deviation derived from region C were scaled by the 
square foot of the ratio of fitted fluxes between regions C and B. Tests made 
for the other quasars confirmed that such estimates between regions B 
and C were consistent with 20\%.
 
In general, the $\zeta$ value increases with heavier smoothing, but individual spectral features become less visible. 
It appears that a smoothing box of 3 pixels is proper in finding spectral voids.
A small fraction of quasars shows low S/N ($< 3$) or contrast ratios ($\zeta < 1.3$). 
They were excluded in this study because of high
uncertainties in determining their proximity zones.

\subsection{Systemic Redshifts}
\label{sec_z}

Accurate quasar redshifts set the zero points of proximity zones and hence are essential
in understanding the level of uncertainties. 
However, this issue remains controversial as emission lines display slightly different redshifts.
The SDSS pipeline provides its
best redshift estimates from multiple emission lines in quasar spectra, which are  
referred to as the SDSS redshifts. \cite{hewett} derived the redshifts of SDSS quasars independently and found a systematic 
difference as a function of redshift.
They suggested that, at $z \approx 3$, \ciii\ may serve as a good reference line. The SDSS spectra after DR 7 provide additional information, including the PCA (Principal Component Analysis) 
redshifts and \ciii\ redshifts. 
As shown in Figure~\ref{fig-dz}, the difference between PCA and pipeline redshifts 
are small: $z(SDSS)-z(PCA)= -0.0010 \pm 0.0053$. The differences for \ciii\ redshifts show larger dispersion:
$z(PCA) - z(C_{III}) = 0.0039 \pm 0.0092$, and
$z(SDSS) - z(C_{III}) =  0.0023 \pm 0.0094 $. Nearly 90\% of the candidates have tabulated \ciii\ redshifts from the SDSS pipeline. For other 
quasars that were processed in DR 7, 
The {\tt Specfit} fitting task was carried out in the wavelength region of $1800 - 1950$~\AA, excluding the 
spectral windows of significant night sky and absorption lines. 
The underlying continuum  was assumed as a power law, and four Gaussian components were used: 
\ion{Al}{3} $\lambda 1857$, \ion{Si}{3}] $\lambda 1892$, and a pair of narrow and broad \ciii\ lines.
The \ciii\ redshifts were derived from the centroids of the fitted narrow \ciii\ components.
When the flux of a narrow \ciii\ component was less than 20\% of the broad 
counterpart, another round of fitting was made with a single Gaussian
 component to determine the \ciii\ centroid.
  
The objects with differences $dz > 0.025$ or $dz< -0.025$
between any pair of the three redshift values were rejected as they are beyond the $2\sigma$ range and may cause considerable uncertainties.
The average difference is $z_{SDSS} - z_{CIII} = 0.0037 \pm 0.0116$ for the samples. If 
the \ciii\ redshifts are used, there would be a systematic shift of \lya\ peak by $-1 \pm 3$~\AA.  
\subsection{Measurement of Proximity Zone}
\label{sec-zone}

As discussed above, some quasar spectra were excluded because of their low quality or unusually large uncertainty in their
redshifts. In addition,  the spectra that display broad absorption lines, which 
account for approximately 12\% of the  sample, were also excluded.
Broad absorption lines make large voids in the RMS-fluctuation spectra that resemble proximity zones. 
Absorption lines in the data are considered  ``broad'' if their EW are larger than $5$~\AA\ between $ 1175 - 1250$~\AA, or 
larger than $ 10$~\AA\ between $ 1050 - 1175$~\AA. All the spectra were visually inspected, and a few spectra with significant artifacts such as flux spikes were excluded. 
Overall, the final sample consists of 2594 quasars. 

Searches for proximity zones were carried out from 1216 \AA\ (the origin) 
towards shorter wavelengths, based on the presence of spectral voids of 3.5~\AA\ or larger in 
RMS-fluctuation spectra.
The task started from the points below a threshold of $\bar{B} + t_0 \zeta \sigma(B)$, where $t_0 = 0.2$.
it then moved towards nearby high points until it hit a threshold $\bar{B} + t_1 \zeta \sigma(B)$, where $t_1 =0.6$. 
The separation between the two endpoints was taken as the void size.
If a void was 3.5~\AA\ or larger, 
it was tentatively selected. In the cases of multiple voids in a quasar's 
vicinity, the shortest wavelength of the furthest void was taken to mark the size of its proximity zone.
The choice of zone-size limit and threshold parameters $t_0$ and $t_1$ affects the estimates
of proximity-zone sizes, and these parameters were optimized over test runs so that the variations of 
zone sizes are the least sensitive to them.
For example, for $t_1$ increased from 0.6 to 0.7, the average zone size of the whole sample becomes 
smaller by 7\%. If it is increased to 0.8, the corresponding change in zone size is 17\%.
In panel 4 of Figure \ref{fig-der}, two spectral voids are found within the radial distance of $2 R_{\omega=1}$:
D ($\sim 1202 - 1206$ \AA) and E ($\sim 1210.5 - 1214$ \AA). Three levels
of $t$ values, 0.2, 0.6, and 1.0, are marked from low to high. 
Since the boundaries of most voids are steep, the $t$ values do not affect the void 
sizes significantly.

The distribution of proximity-zone sizes is plotted in Figure \ref{fig-zone}.
The bin with the largest number of proximity-zone sizes
is at $0-5$ Mpc, and the numbers decline towards larger distances
exponentially with a correlation coefficient $ r = -0.99$ between
$R$ and $\log(N)$ and an e-folding distance of 3.6 Mpc.
Since proximity zones are dependent on quasar luminosities, it is useful to normalize the 
zone sizes by quasar luminosities.
Characteristic sizes, $R_{\omega=1}$,
were calculated from the $i'$-band magnitude and nominal optical-UV and 
extreme-UV (EUV) spectral shapes. 
They serve only as a coarse scale because the quasar's ionizing continuum is 
extrapolated using an average EUV power law \citep{zheng97,telfer}, and the UVB is fluctuating \citep{meiksin19}. 
The normalized proximity-zone sizes, namely
 the ratios of measured to characteristic values, 
also distribute exponentially ($r = -0.99$) with an e-folding scale of 0.64.

\section{RESULTS}\label{sec_res}

\subsection{Quasars with Large Proximity Zone} \label{sec_large}

A group of quasars is selected for their large proximity zones, based on the following criteria: (1) a spectral void that starts (at its longest 
wavelength) between 1.0 and 2.0  $R_{\omega=1}$ and with a size of $\geq 3.5 $ \AA, (2) a minimum zone size of 10 Mpc, as measured at the ending (the shortest) wavelength, 
and (3) no absorption lines at $z_{ab} > z_{em}$.
Ninety-two quasars ($\sim 4$ \% of the sample) match these criteria. 
They are potentially old quasars because of their large ionized zones, but see 
\S \ref{sec_age} for more justifications.

A composite spectrum with a pixel scale of 1.0~\AA\ was generated for 
this group. 
All spectra were first normalized around 1350~\AA. 
For every merging spectrum, the pixels that fall into the absorption windows were 
flagged and excluded. Within 20 \AA\ from the \lya\ wavelength, these flagged data 
points were filled with fitted values (see \S \ref{sec_prep}). 
Every data point in the composite 
spectrum took the {\em median} of the fluxes of merging spectra at this wavelength.
As shown in the lower panel of Figure~\ref{fig-comp}, 
the spectrum of this group shows strong narrow components in major emission lines, most notably \lya.
Several other combining algorithms were also tried, such as an equal weight for every 
source and an average value at each wavelength bin. While the resultant spectra 
differ in details, a strong narrow \lya\ core is always present.

\subsection{Quasars with Infalling System and No Proximity Zone} \label{sec_small}

Since approximately half of the quasars in the sample display small or no proximity
zones,  it is necessary to find a subgroup among them that is most 
representative of young quasars. 
It is noted that some quasars in this sample display narrow absorption features 
on the red wing of \lya\ emission lines, and nearly all them ($>95$\%) exhibit 
no proximity zones. 
Another group is therefore selected based on the lack of proximity zones (no spectral voids of 
$\geq 3.5$ \AA\ and no voids of $\geq 2$ \AA\ within 3.5 \AA\ from \lya) and the 
presence of infalling systems at $z_{ab} > z_{em}$:
(1) The absorption redshift is larger than any of the tabulated \lya\ redshifts $-$
the SDSS pipeline redshift, \ciii\ redshift 
and PCA redshift. The longest \lya\ wavelength among these three is considered as the lower end 
of the selection window, which extends to 1226~\AA. (2) Only significant absorption features 
with EW $> 0.5$~\AA\ are selected. (3) Potential doublets of metal absorption, 
including \nv, \civ, and \mgii, are rejected. These features are flagged when their wavelength 
separations are within 0.5~\AA\ from the expected theoretical values, their intensity 
ratios within 35\% of the theoretical value ($0.5-1.0$). 

These 77 objects, which account for approximately 3\% of the sample, are characterized by 
weak narrow cores of 
emission lines. While it is possible that strong absorption features near the \lya\ wavelength 
may hinder the narrow \lya\ core, the effect is mitigated using the fitted values in these 
flagged data points of each spectrum.

The upper panel of Figure~\ref{fig-comp} shows the composite spectrum for this group of potentially young quasars. The most significant feature for these young quasars is the weak narrow core in \lya.
In the middle panel, the composite spectrum for all the quasars in this sample is plotted.

To demonstrate the sharp difference in emission-line profiles, Figure~\ref{fig-exam} displays the spectra of two individual quasars  
from these groups, along with their RMS fluctuations. The quasar spectrum
in the top left panel displays a strong narrow \lya\ component. In the lower
 left panel, several spectral voids suggest a proximity zone as large as 
24 \AA\ ($\sim 20$ Mpc).
In the top right panel, another quasar spectrum displays a broad \lya\ profile and no proximity zone. 

\subsection{Comparison of Spectral Properties}
\label{sec_fit}

The {\tt Specfit} task was carried out for the three composite spectra 
of young quasars, old quasars, and the full sample.
The fitting components included a broken power law, Gaussian profiles for emission lines and 
a set of smooth absorption profile with different optical depths shortward of the \lya\ wavelength.
For strong emission lines, dual components were used: A narrow Gaussian of FWHM$\sim 2300$ \kms, and
a broad one of $\sim 11000$ \kms. The widths for \ovi, \nv, \civ, and \ciii\ 
components were tied to 
those of \lya. For weak lines of \oi, \cii, and \siiv, single Gaussian components
were used.
As shown in Table \ref{tbl-comp}, the main difference between the young and old quasars are the
EW of narrow-line components, with the EW of narrow \lya\ varying by a factor of ten. 
The EW of broad components are similar in all three groups, with relative differences 
of $\lesssim 30\%$. 
Table \ref{tbl-width} lists the FWHM of the fitted emission lines.

The UV continua of young quasars appear slightly redder than the old ones. The breaking wavelength
 for the dual power-law components is near \lya, and
the average power-law indices ($f_\lambda \propto \lambda^{-\beta}$) at wavelengths longward 
of \lya\ are $\beta = 1.24 \pm 0.37$ 
for young quasars, $1.46\pm 0.32 $ for old quasars and $1.38 \pm 0.41$ for the whole sample.
The fitting results of the composite spectra yield $\beta = 1.22 \pm 0.01$ for the young 
quasars, $ 1.56 \pm 0.02$ for the old ones, and $1.47 \pm 0.01$ for the whole 
sample. 

The three groups have similar luminosities: the average absolute magnitudes are $-27.79 \pm 0.39$ for 
the young quasars, $-27.82 \pm 0.41$ for the old quasars, and $-27.81 \pm 0.44$ for the whole 
sample. The average redshift for the young quasars is $2.86 \pm 0.29$, slightly higher than that
of the old quasars ($2.80 \pm 0.26$). The observed difference in emission lines likely reflects an evolution 
trend over a quasar's lifetime instead of a luminosity or redshift effect.

To estimate the sample dispersion, 100 composite spectra 
were generated by randomly selecting 
subgroups of 50 quasars from a parent sample. 
A dispersion spectrum was generated by taking the 
standard deviations at each wavelength bin. As shown in Figure \ref{fig-disp}, the intrinsic 
dispersions are similar for different groups, with larger dispersion at emission-line 
regions and the \lya\ forest region.

\section{DISCUSSION}

\subsection{Effect of Light Travel Time}
\label{sec_age}

The observed expansion of \h\ proximity zones along lines of sight
is believed to be superluminal
because an ionizing front propagates at nearly the speed of light in the quasar vicinity 
\citep{white}. Several IGM models assumed an infinite speed for the 
line-of-sight development of a proximity zone \citep{bolton07,lu,khrykin}, thus 
ruling out the possibility of using the \h\ proximity effect to scale quasar ages. This 
hypothesis can be tested with \he\ proximity profiles, as their development should display 
significant lags with respect to their \h\ counterparts. \cite{zheng19} studied the \he\ and \h\ 
proximity zones 
in 15 quasars and found a significant correlation between them. Since the sizes of \he\ proximity 
zones are believed to be related to quasar ages \citep{khrykin,khrykin19}, the \h\ zone sizes 
should also bear the signature of quasar ages. The correlation between the sizes of \he\ and \h\
proximity zones suggests that the expansion of quasar ionizing fronts may be noticeably slower than the 
speed of light. One possibility is that quasar activities are episodic on a time scale of the IGM's
equilibration ($\approx 10^4$ yr).
While more work is needed to explain this observed trend, the relation may serve 
as an empirical gauge to probe young and old quasars from the sizes of their \h\ proximity zones. 

\subsection{Effect of External Sources} \label{sec_ext}

Quasar sightlines pass through a vast volume of the IGM, whose properties are subject to significant 
variations. The forest-line spectra of some quasars display significant voids 
\citep{dobrzycki,syphers14}, which are believed to be the IGM 
clustering properties or the transverse proximity effect due to the foreground quasars near sightlines. 
To test the level of contamination by external sources, the distribution of spectral voids is studied at different wavelengths.  
As shown in Table \ref{tbl-void} and Figure \ref{fig-void}, the majority of quasars display spectral voids of 3.5 \AA\ and larger. 
The probability of finding such voids at wavelengths below 1170 \AA\ is 
$8.8 \pm 0.9\%$ per interval of 10~\AA. This percentage
increases significantly to $\sim 46 \%$ in the wavelength range of $1205 - 1215$~\AA,
providing evidence that the majority of voids in the quasar vicinity are 
associated with the intrinsic proximity effect.
The contamination rate is estimated at approximately 20\%, but the actual 
rate is lower as the voids with their RMS-fluctuation troughs lower than those at longer 
wavelengths, a common feature for external spectral voids, were rejected.

\subsection{Effect of Overdensity}
\label{sec_dens}

It is common to assume that the dark matter in the quasar vicinity is denser than 
the IGM's average, as quasars are believed to be associated with the most massive galaxies 
\citep{kauffmann,springel}. The net effect for a higher density is a higher number of absorbers 
and smaller proximity zones.
\cite{guimaraes} found that the mean overdensity is of the order of two and five within, 
respectively, 10 and 3 Mpc from a quasar. At such levels of overdensity, the characteristic sizes 
would be reduced modestly, but most proximity zones would not be dramatically reduced or masked.
It would be ever harder to explain the proximity zones larger than 10 Mpc in terms of 
underdensity, as the IGM density would have to be lower by a factor of $\sim 4-10$ in a large volume 
along these lines of sight.

\subsection{Effect of Viewing Angles}

The differences in line profiles between the young and old quasars bear resemblance to that between the Type-I and II active galactic nuclei (AGNs). 
The narrow \lya\ cores in the group of old quasars are of FWHM $\approx 1500-3500$~\kms. 
As a comparison, nearly all the Type-II SDSS quasars display line widths FWHM(\hb)$<1500$ 
\kms\ \citep{zakamska}, and a new study has set an upper limit of 1000 \kms\ \citep{yuan}.
These \lya\ emission profiles also display significant broad 
wings; therefore, the sample consists of Type-I quasars only. 
But would it be possible that the observed differences are attributed to viewing angles?

Quasar radiation is believed to be highly anisotropic, therefore, the surrounding 
proximity zone is not spherical.
According to the unification theory of AGN \citep{urry}, Type-I AGN display broad emission lines 
as they are viewed at head-on directions towards obscuring tori. Viewing angles affect 
the observed broad-line widths and EW \citep{rudge}. 
The quasar radiation is expected to be strong at such viewing angles,
therefore, larger proximity zones and broader emission lines are anticipated. This seems
inconsistent with the results shown in Figure \ref{fig-comp}. 

\cite{fine} studied the radio spectral indices of SDSS quasars and found broader \mgii\ emission profiles
for quasars with steep spectral indices, suggesting disk-like velocities for the broad-line region (BLR). 
The results in Table \ref{tbl-comp} are consistent with a model of two components in the BLR: 
intermediate- and very-broad-line regions \citep{brotherton,hu,zhu}. If their relative strengths or the 
sizes of inner tori \citep{simpson,elitzur} vary during a quasar's lifetime, it might explain the 
differences in broad emission-line profiles as discussed above.

\subsection{Effect of Analysis Parameters} \label{sec_par}

The changes in spectral features shown in \S \ref{sec_res} 
reflect a gradual trend over proximity-zone sizes, and the results in the 
moderate zone sizes are dependent on the selection parameters. 
The current results are based on thresholds $t_0 = 0.2$, $t_1 = 0.6$, 
and a minimum void size of 3.5 \AA. 
These parameters were selected from many test runs. As shown in 
Table \ref{tbl-void}, the number of spectral voids decreases with higher size limits, reducing the contamination from external voids along lines of sight. 
For a nominal void-size limit of 3.5 \AA, the contamination level is $\sim 20\%$.
This ratio decreases to $\sim 13\%$ for a higher limit of 4.5 \AA, and increases to 
$\sim 33\%$ for a lower limit of 2.5 \AA. However, too high a void-size limit
would result in missing proximity zones. Several quasars in the sample 
have been studied with their Keck high-resolution spectra \citep{zheng19}, 
and a comparison with the proximity measurements found a reasonable match with
the current selection parameters.

Figure \ref{fig-par} illustrates the effect on \lya\ profiles 
when selection parameters vary. 
For a void-size limit of 2.5 \AA, as shown in the green curve, 
the flux of \lya\ peak is lower by approximately 10\% for the composite spectrum of old quasars.
When the limit is set to 4.5 \AA, the red dashed curve is almost 
indistinguishable from that of 3.5 \AA. 
This is because the majority of selected old quasars display void sizes greater than 5 \AA.  
Indeed, the choice of void-size limits or the threshold levels would affect the histogram 
distribution in Figure \ref{fig-zone}. Higher $t_1$ values would lead to larger 
voids. As the red dotted curve in Figure \ref{fig-par} shows, the differences 
are small. If $t_0$ is set to zero, more voids may be identified, but it makes
no difference for the subgroup of old quasars.

A test is also made to take the average flux values at each wavelength bin in the composite spectrum. 
As shown in the blue curve in Figure \ref{fig-par}, the difference between the median and mean values is 
less than 10\%.  The characteristics of composite spectra of old and young quasars are therefore not just 
the results of a specific data analysis with certain parameters. 

\section{CONCLUSION}\label{sec_conc}

Based on the spectral voids between \lya\ forest lines in a large sample of 
2594 SDSS spectra of quasars,
the proximity zones are measured to distances considerably larger than previous studies.
The majority of quasars display small zones of  $\lesssim 5$~Mpc 
in proper distance.
A group of old quasars are identified with large proximity zones. They display strong narrow cores of emission lines.
Another group of young quasars show infalling components ($z_{ab} > z_{em}$) and no 
proximity zones. The narrow \lya\ components in these young quasars are weak. 
The ratio of narrow and broad components in \lya\ increases from 0.05 for young quasars to 0.5 
for old quasars. 

The differences in line intensities and profiles have been a focus of quasars studies, as they are 
among the main PCA drivers  \citep{boroson}. 
\cite{suzuki} carried out a PCA using 50 quasar spectra at lower redshifts and suggested four classes of different 
emission-line properties. Interestingly, the Class I objects in his work bear resemblance to
old quasars  and the Class III are similar to young quasars. The underlying reasons for such differences 
remain to be explored, and they could be related to quasar ages.
In the early stage of a quasar's evolution, the supermassive halo around it is rich in infalling systems. It is also possible that 
broad absorption 
lines are common at this stage. Old quasars, on the other hand, may be in an
evolutionary stage that lacks infalling matter. 

Ample evidence suggests that different line widths and EW 
are the results of viewing angles. The results presented in this paper suggest
that they may also be related to quasar ages. Further studies of the broadband properties 
of these groups may reveal more evolution effects and shed light on the underlying 
reasons of known quasar properties. 
Since the strength of narrow cores of emission lines evolves with the sizes of proximity zones, 
it may serve as an age indicator at lower redshifts, when the information of forest lines is 
not readily available.
Several known properties of emission lines, such as the Baldwin effect, may be understood in terms of 
quasar ages: young quasars show broad emission profiles, and narrow core become stronger in later 
stages as the Eddington ratio declines, and the BLR configuration changes.

\acknowledgments 

The author thanks H. C. Ford, G. A. Kriss, Y. Lu,  A. Meiksin, N. L. Zakamska, and an anonymous 
referee for many constructive comments.

Funding for the SDSS, SDSS-II, and SDSS-III has been provided by the Alfred P. Sloan Foundation, 
the Participating Institutions, the National Science Foundation, the U.S. Department of Energy, 
the National Aeronautics and Space Administration, the Japanese Monbukagakusho, the Max Planck Society,
and the Higher Education Funding Council for England. The SDSS Web Site is http://www.sdss.org/.

The SDSS is managed by the Astrophysical Research Consortium for the Participating Institutions. 
The Participating Institutions of the SDSS-I-II-III include American Museum of Natural History, 
University of Arizona, 
Astrophysical Institute Potsdam, 
University of Basel, 
the Brazilian Participation Group, 
Brookhaven National Laboratory, 
University of Cambridge, 
Carnegie Mellon University, 
Case Western Reserve University, 
University of Chicago, 
Drexel University, 
Fermilab,
University of Florida, 
the French Participation Group, 
the German Participation Group, 
Harvard University, 
Institute for Advanced Study, 
Instituto de Astrofisica de Canarias, 
the Japan Participation Group, 
Johns Hopkins University, 
Joint Institute for Nuclear Astrophysics, 
Kavli Institute for Particle Astrophysics and Cosmology, 
the Korean Scientist Group, 
the Chinese Academy of Sciences (LAMOST), 
Lawrence Berkeley National Laboratory, 
Los Alamos National Laboratory, 
Max Planck Institute for Astronomy, 
Max Planck Institute for Astrophysics, 
Max Planck Institute for Extraterrestrial Physics, 
the Michigan State/Notre Dame/JINA Participation Group, 
New Mexico State University,  
New York University, 
Ohio State University, 
Pennsylvania State University, 
University of Pittsburgh, 
University of Portsmouth, 
Princeton University, 
the Spanish Participation Group, 
University of Tokyo, 
United States Naval Observatory, 
University of Utah, 
Vanderbilt University, 
University of Virginia, 
University of Washington, 
and 
Yale University.

\setcounter{figure}{0}
\begin{figure}
\plotone{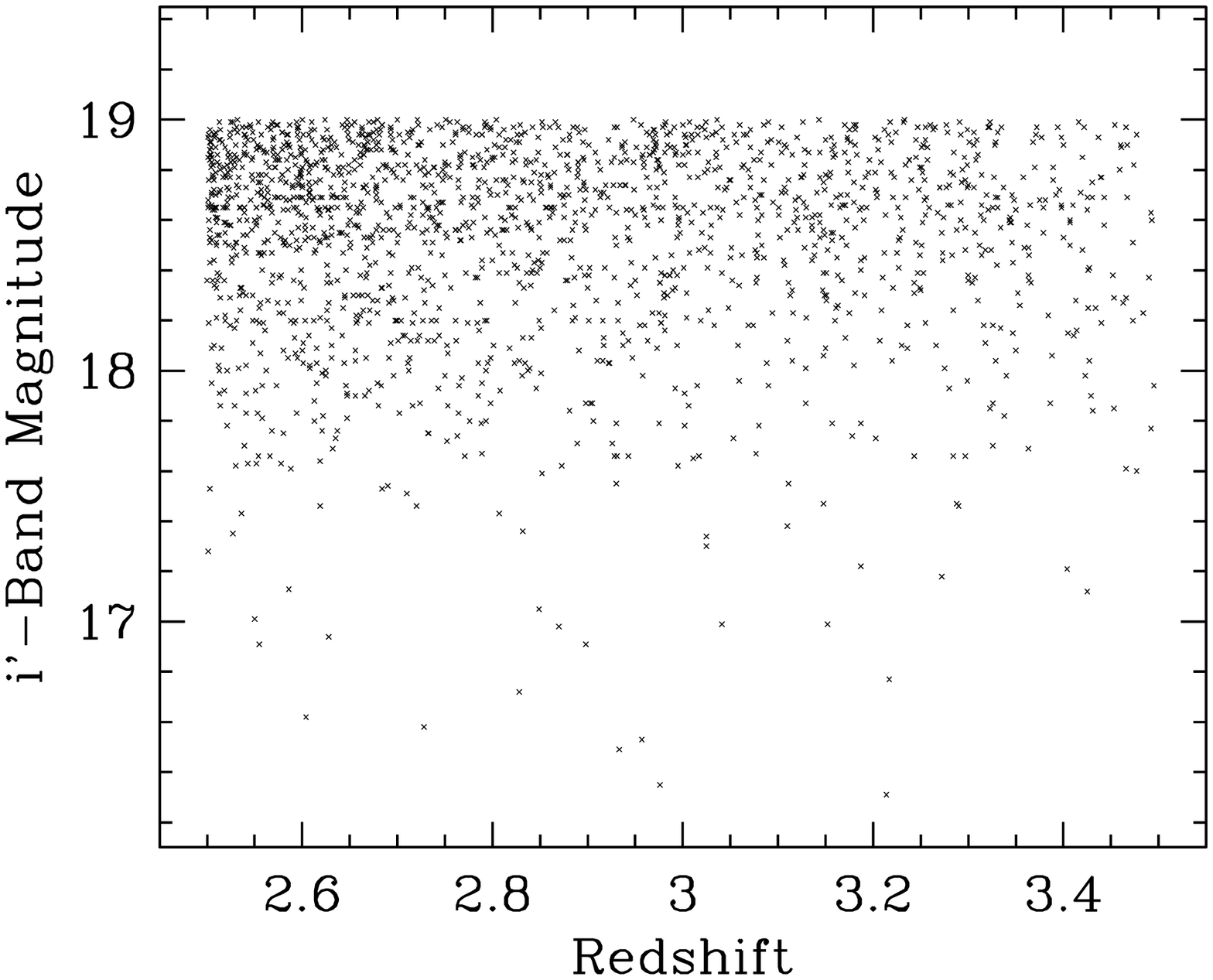}
\caption{SDSS quasars at $2.5 \leq z \leq 3.5$.  Only  2594 of these 3298 spectra are actually used in the analysis, 
as others are rejected for low quality, large redshift uncertainties, or broad absorption features.
\label{fig-sam}}
\end{figure} 

\begin{figure}
\plotone{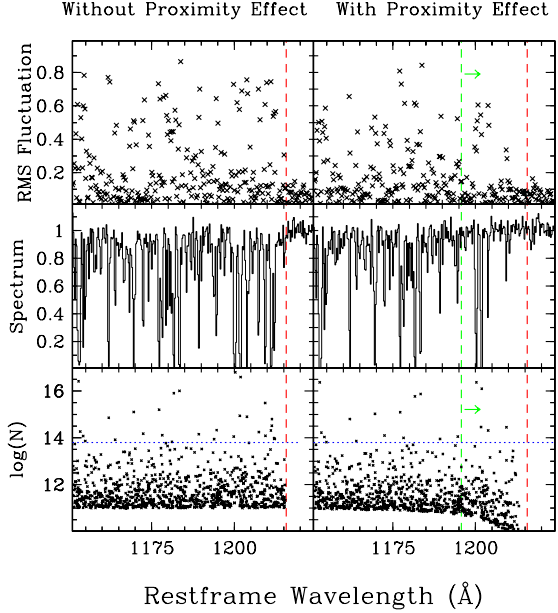}
\figcaption{Simulations of \lya\ forest lines at $z=3$. The left panels show 
the original properties without the proximity effect, and the right panels 
with a proximity zone of $R_{\omega=1} = 20$~\AA, as marked with a green dashed line  
and arrows. The red dashed lines mark the \lya\ wavelength.
The lower panels are the column density $N$ of simulated absorbers, in units of \cl,
the middle panels the forest spectra, and the upper panels the RMS-fluctuation spectra.
The spectral voids near \lya\ in the right panels are 
associated with absorbers of 
column density $\log N \lesssim 13.8$~\cl\ below the blue dotted line. 
\label{fig-sim}}
\end{figure} 

\begin{figure}
\plotone{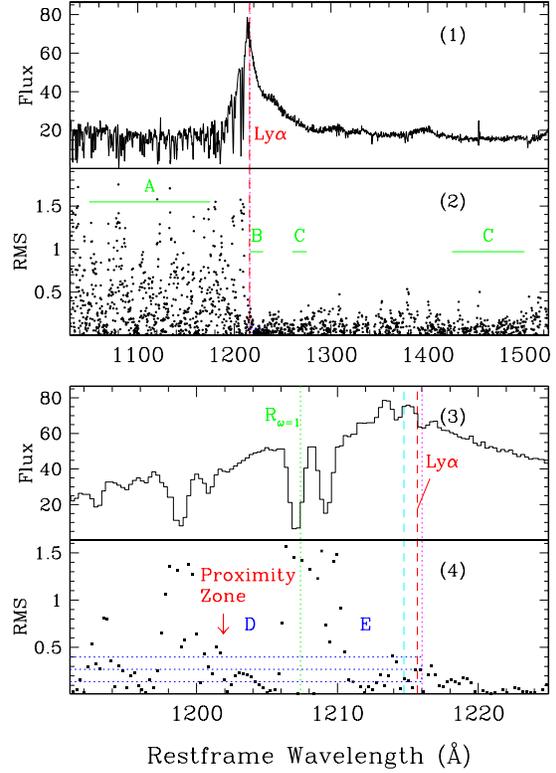}
\figcaption{Spectrum of quasar SDSS1438+0831, $z=2.839$, and its RMS fluctuations.
The flux unit is $10^{-17}$ \ergscmA, and the RMS spectra are smoothed by 3 pixels. 
In panel 2, the average value in region A is 0.40, and $0.074 \pm 0.057$ in region B, yielding $\zeta\approx 5.4$. 
Region C is marked but not used. 
In panels 3 and 4, the \lya\ wavelength is marked in red, that for the PCA 
redshift in magenta dots, that for the \ciii\ redshift in cyan, and the characteristic 
distance of $R_{\omega=1}$ in green. Three blue horizontal dotted lines 
represent the levels of $t=0.2$, 0.6, and 1.0, respectively, from low to high. There 
are two voids, D and E, and the proximity-zone size is marked with a downward arrow at the left boundary of D.
\label{fig-der}}
\end{figure} 

\begin{figure}
\plotone{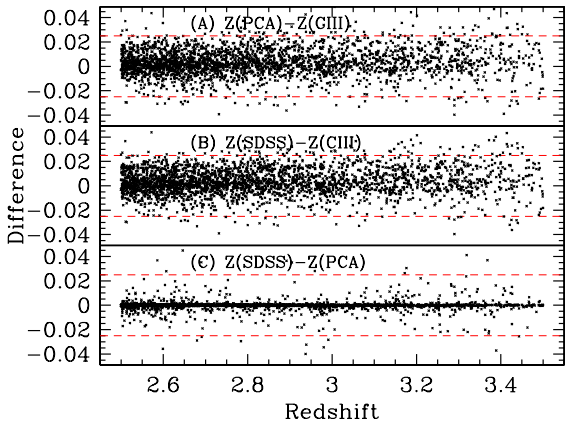}
\figcaption{
Redshift differences between the three sets of SDSS measurements. Panel A: between the PCA redshifts
and the \ciii\ redshifts; panel B: between SDSS pipeline redshifts, and \ciii\ 
redshifts, with $z(SDSS) - z(C_{III}) =  0.0023 \pm 0.0094 $; and panel C: between the SDSS pipeline redshifts, and PCA redshifts.
Objects beyond the range $-0.025 < dz < 0.025$, as marked with red dashed lines, are not used in the study.
\label{fig-dz}}
\end{figure} 

\begin{figure}
\plotone{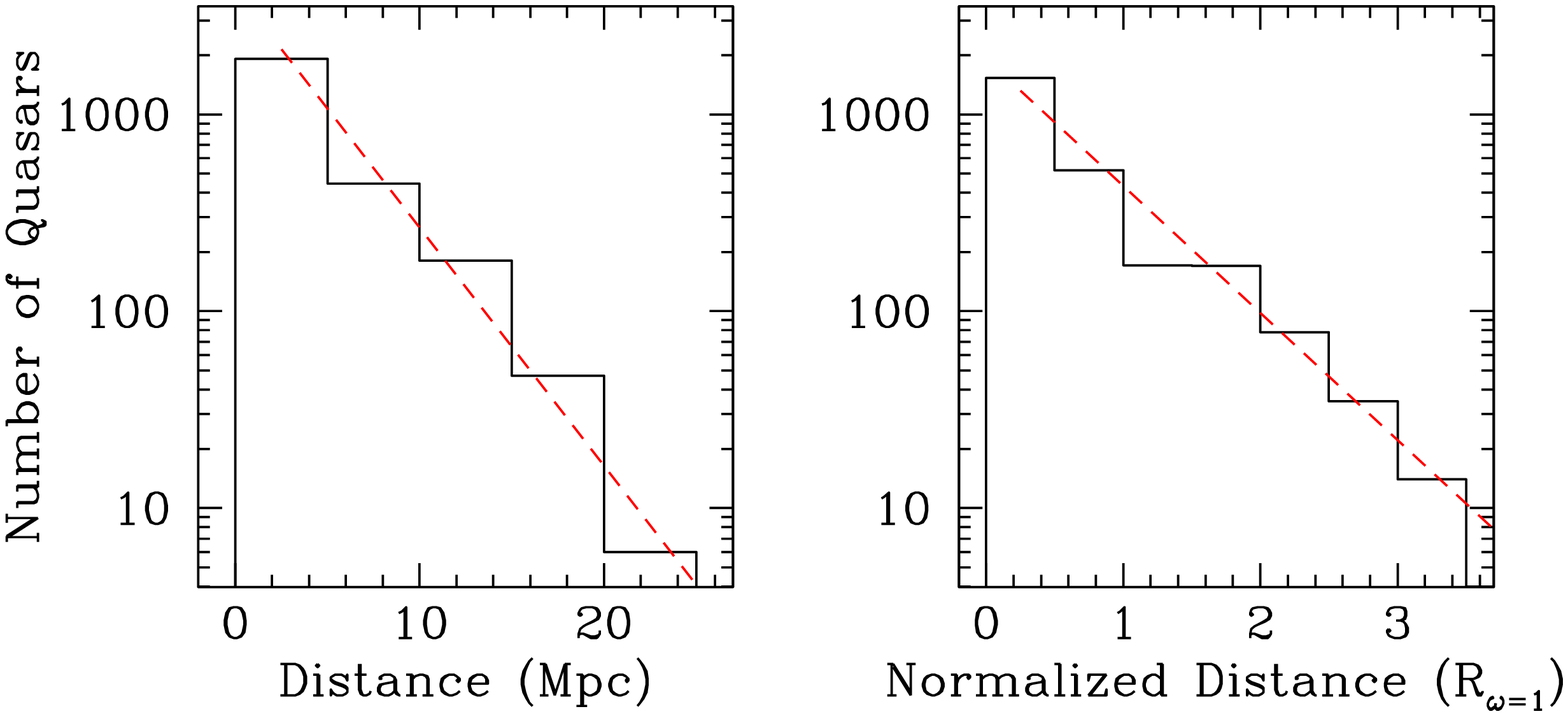}
\caption{
Distribution of proximity zones. In the right panel, proximity zones 
are normalized to the characteristic sizes $R_{\omega=1}$.
A zone size is defined by the distance between the \lya\ wavelength and the furthest data point in 
a spectral void. The samples of young and old quasars are subgroups selected with additional criteria (see \S 
\ref{sec_res}).
Both distributions can be fitted exponentially with 
high confidence (red lines).
\label{fig-zone}}
\end{figure}

\begin{figure}
\plotone{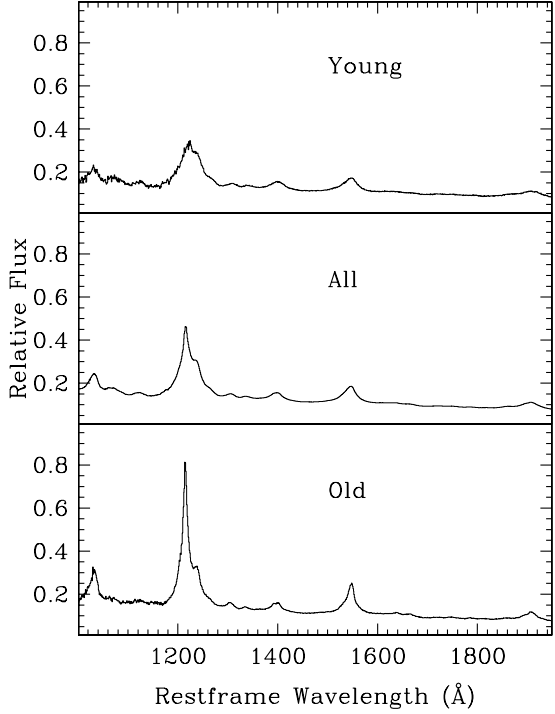}
\caption{
Three composite quasar spectra with different proximity properties 
and implied ages. 
The top panel is for the young quasars with no proximity zones and 
with significant absorption lines longward of the \lya\ wavelength, the middle panel for all 
the quasars in the sample, and the bottom panel the old quasars 
with large proximity 
zones ($> 10$~Mpc). The spectra are scaled to display similar flux levels around 1350~\AA.
\label{fig-comp}}
\end{figure} 

\begin{figure}
\plotone{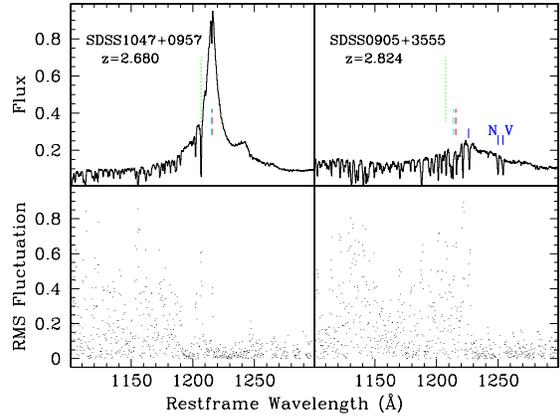}
\caption{
Spectra of two distinct quasars and their RMS-fluctuations. Top left 
panel: a spectrum with a large proximity zone down to $\sim 1192$ \AA.
Top right panels: another spectrum with no proximity zone. 
A pair of \ion{N}{5} $\lambda\lambda 1238/1242$ and associated 
\lya\ absorption lines are marked in blue as an infalling system 
at $z_{ab} = 2.858$. Red and cyan bars mark the \lya\ wavelengths derived from the SDSS 
redshifts, which are identical to the PCA values in most cases, and the \ciii\ redshifts, respectively. 
The long green dotted lines mark the respective characteristic zones 
$R_{\omega=1}$.
The two spectra are scaled so that their fluxes are equal around 1300~\AA. 
The two RMS-fluctuation spectra are roughly normalized.
\label{fig-exam}}
\end{figure} 

\begin{figure}
\plotone{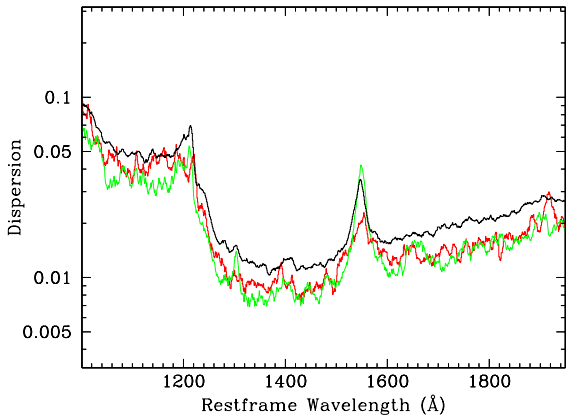}
\caption{Dispersion spectra as a fraction of the respective composite spectra. 
Black curve: all quasars; red curve: young quasars, and green curve old quasars. 
\label{fig-disp}}
\end{figure} 

\begin{figure}
\plotone{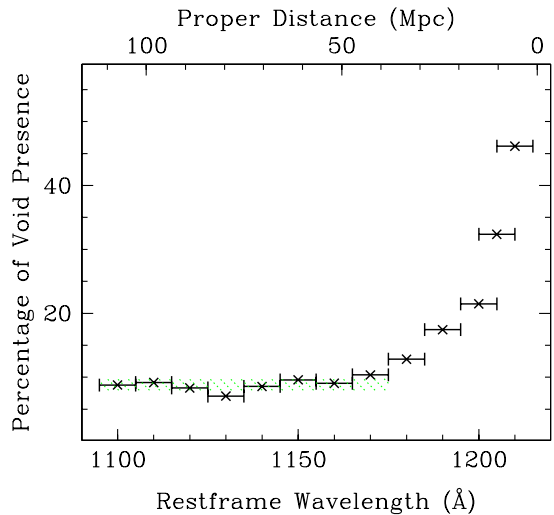}
\caption{
Number of spectral voids of size 3.5~\AA\ or larger in the rest frame. 
The search is carried out in the RMS-fluctuation spectra of 2594 quasars. The green shaded region marks the $1\sigma$ confidence range for the results 
between 1095 and 1175~\AA. 
The top labels mark the distances to a quasar at $z=2.8$. 
The proximity effect is present beyond $10\sigma$ at wavelengths longer than 
1190~\AA, or radial distances up to $\sim 20$~Mpc.
\label{fig-void}}
\end{figure} 

\begin{figure}
\plotone{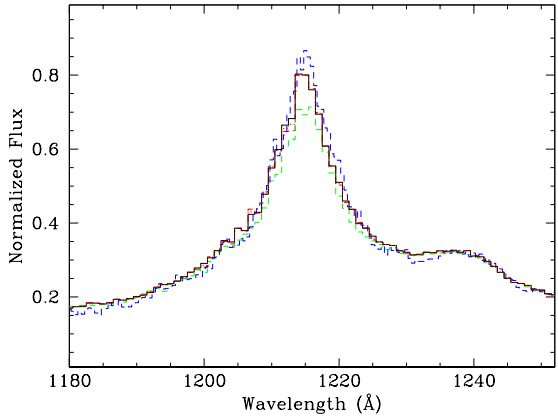}
\caption{
\lya\ profiles of old quasars produced with different selection parameters. The black 
curve: 
the composite spectrum of current selection, as shown in the lower panel of Figure \ref{fig-comp}, which 
is made of median values at each wavelength bin; the green curve: with a minimum void size of 2.5 \AA; 
the red dashed curve: with a minimum void size of 4.5 \AA; and the red dotted
curve: $t_1=1.0$. No difference is found between $t_0=0.2$ and 0.
The blue curve is made with mean values at each pixel, using the current selection.
\label{fig-par}}
\end{figure} 

\begin{deluxetable}{ccccccc}
\tablecaption{Equivalent Widths of Fitted Emission Lines in Composite Spectra\tablenotemark{a}\label{tbl-comp}}
\tablewidth{0pt}
\footnotesize
\tighttable
\tablehead{
\colhead{Line} &
\multicolumn{2}{c}{Young} &
\multicolumn{2}{c}{All} &
\multicolumn{2}{c}{Old}
\\ \colhead{} &
\colhead{Broad} &
\colhead{Narrow} &
\colhead{Broad} &
\colhead{Narrow} &
\colhead{Broad} &
\colhead{Narrow} 
}
\startdata 
\ovi & $12.6\pm 3.6$ & $ 1.5 \pm 0.8$ &$13.4 \pm 0.3$ &$2.0 \pm 0.1$& $17.9\pm 0.7$ &$3.4 \pm 0.2$ \\
\lya & $50.0 \pm 1.7$ & $0.4 \pm 1.0$ &$53.9 \pm 0.3$&$9.3 \pm 0.2$&$51.8\pm 1.3$ &$27.1 \pm 0.9$ \\ 
\nv & $17.0\pm 2.0 $&$1.5\pm 0.4$ &$10.9 \pm 0.2$ &$2.4 \pm 0.1$&$17.2 \pm 2.3$ &$3.4 \pm 0.2$  \\ 
\civ &$21.6 \pm 1.0$ & $3.5 \pm 0.9$ &$23.8 \pm 0.3$&$5.4\pm 0.1$&$26.0\pm 0.7$ &$11.3\pm 0.4$ \\ 
\ciii\ $\lambda 1909$ & $18.5 \pm 1.2$ &$1.7 \pm 0.8$ &$19.8 \pm 0.4$&$3.0 \pm 0.3$& $18.1\pm 1.7$ & $4.7 \pm 0.6$  \\ 
\oi & \multicolumn{2}{c}{$2.1 \pm 0.2$} & \multicolumn{2}{c}{$1.7 \pm 0.1$} & \multicolumn{2}{c}{$2.4 \pm 0.2$} \\
\cii & \multicolumn{2}{c}{$1.6 \pm 0.3$} & \multicolumn{2}{c}{$0.7 \pm 0.1$} & \multicolumn{2}{c}{$0.5 \pm 0.1$} \\
\siiv & \multicolumn{2}{c}{$8.9 \pm 0.3$} & \multicolumn{2}{c}{$7.3 \pm 0.2$} & \multicolumn{2}{c}{$6.5 \pm 0.3$} \\ \enddata 
\tablenotetext{a}{in units of \AA. The errors are the {\tt Specfit} fitting results 
and do not include those between individual spectra.}
\end{deluxetable}

\begin{deluxetable}{ccccccc}
\tablecaption{Full Widths at Half Maximum
of Fitted Emission Lines in Composite Spectra\label{tbl-width}}
\tablewidth{0pt}
\footnotesize
\tighttable
\tablehead{
\colhead{Line} &
\multicolumn{2}{c}{Young} &
\multicolumn{2}{c}{All} &
\multicolumn{2}{c}{Old}
\\ \colhead{} &
\colhead{Broad} &
\colhead{Narrow} &
\colhead{Broad} &
\colhead{Narrow} &
\colhead{Broad} &
\colhead{Narrow} 
}
\startdata 
Major Lines\tablenotemark{a}& $10915 \pm 480$ & $2331 \pm 477$ &$11155 \pm 75$&$2498 \pm 37$& $10002\pm 238$ &$ 2429 \pm 48$ \\ 
\oi & \multicolumn{2}{c}{$4080 \pm 387$} & \multicolumn{2}{c}{$3466 \pm 224$} & \multicolumn{2}{c}{$3362 \pm 274$} \\
\cii & \multicolumn{2}{c}{$4046 \pm 1093$} & \multicolumn{2}{c}{$2726 \pm 405$} & \multicolumn{2}{c}{$1797 \pm 521$} \\
\siiv & \multicolumn{2}{c}{$6783 \pm 239$} & \multicolumn{2}{c}{$5761 \pm 151$} & \multicolumn{2}{c}{$4722 \pm 223$} \\ \enddata 
\tablenotetext{a}{For \lya, \ovi, \civ\ and \ciii}
\end{deluxetable}

\begin{deluxetable}{cc}
\tablecaption{Number of Spectral Voids\tablenotemark{a}\label{tbl-void}}
\tablewidth{0pt}
\tighttable
\tablehead{
\colhead{Void Size (\AA)} &
\colhead{Number of Voids} }
\startdata 
$\geq 3$  & 2217   \\ $\geq 3.5$  & 1420 \\    
$\geq 4$  & 936 \\    
$\geq 5$  & 375 \\    
$\geq 6$  & 157 \\   
$\geq 7$  & 82  \\    
$\geq 8$  & 44 \\    
$\geq 10$ & 18 \\    
\enddata
\tablenotetext{a}{Counted from 2594 quasar spectra between 1095 and 1175~\AA. 
Some spectra display multiple voids.}
\end{deluxetable}
\clearpage 

\end{document}